\documentclass{article}

\usepackage{amsmath}
\usepackage{amssymb}
\usepackage{enumerate}
\usepackage{mathrsfs}
\usepackage{color}

\usepackage{lipsum}

\newtheorem{theorem}{Theorem}[section]

\newtheorem{definition}[theorem]{Definition}

\newtheorem{conjecture}[theorem]{Conjecture}

\newtheorem{remark}[theorem]{Remark}

\numberwithin{equation}{section}

\def\R{{\mathbb{R}}}

\def\N{{\mathbb{N}}}
\def\x{{\mathbf{x}}}
\def\G{{{G}}}
\def\a{{\boldsymbol{\alpha}}}
\def\b{{\boldsymbol{\beta}}}
\def\A{{\mathscr{A}}}
\def\B{{\mathscr{B}}}
\def\Conv{{\hbox{\rm{conv}}}}
\def\supp{{\hbox{\rm{supp}}}}
\def\bo{{\boldsymbol{\omega}}}
\def\RGA{{\textrm{RGA}}}
\def\IRGA{{\textrm{IRGA}}}

\begin{document}

\title{On Radically Expanding the Landscape of\\ Potential Applications for\\ Automated Proof Methods}
\author{Jeffrey Uhlmann\thanks{(Corresponding author) 201 Naka Hall, Dept.\ of Electrical Engineering and Computer Science, University of Missouri-Columbia (uhlmannj@missouri.edu, 573.884.2129)} ~and 
Jie Wang\thanks{School of Mathematical Sciences, Peking University (wangjie212@mails.ucas.ac.cn)}}
\date{}
\maketitle

\vspace{-24pt}
\begin{abstract}
In this paper we examine the potential of optimization-based computer-assisted proof methods to be applied much more widely than commonly recognized by engineers and computer scientists. More specifically, we contend that there are vast opportunities to derive valuable mathematical results and properties that may be narrow in scope, such as in highly-specialized engineering control applications, that are presently overlooked because they have characteristics atypical of those that are conventionally studied in the areas of pure and applied mathematics. As a concrete example, we demonstrate use of sum-of-squares (SOS) optimization for certifying polynomial nonnegativity as a part of a proposed dimension-pinning strategy to prove that the inverse of the relative gain array (RGA) of a $d$-dimensional positive-definite matrix is doubly-stochastic for $d\leq 4$. However, it is not specifically this result and solution method that are of principal interest in this paper but rather how they illustrate the relevance of optimization-based proof techniques to engineering system design more broadly.\\
~\\
{\bf Keywords}: Automated Proofs, Automated Software Methods, Computer-Aided System Design, Computer-Assisted Proofs, 
SOS Proofs, Sum of Squares Optimization.
\end{abstract}

\section{Introduction}

Many practical applied mathematics and engineering disciplines, such as control theory, include problems that are computationally intractable 
or even insoluble when formulated in their full generality. However, practical problem solving not just about finding highly general methodologies, it is about deriving specialized solutions tailored to the properties and constraints
of a particular problem of interest. For example, if a system can be shown to be effectively approximated using 
a linear model then a vast array of solution techniques and tools immediately become available. In other cases 
simply knowing that the value of a particular input parameter is bounded within a known interval, or that the 
value of a particular variable cannot exceed a known threshold, may be enough to permit 
-- {\em possibly after prodigious effort and analysis} -- a practical solution to be derived that has
significantly better and rigorously provable 
performance properties. It is often the case, however, that empirical analysis of a system reveals behavior
that is suggestive of fundamental properties that could be productively exploited if formally proven. 
Unfortunately, the types of problems that naturally tend to arise in practical  applications
require proof methods that are typically far removed from the focus of conventional mathematics. 

The goal of this paper is to make a case that many optimization tools and techniques presently exist that 
are not widely recognized but that can aid in the establishing and formal proving of useful
system properties. Such established properties may not only improve performance but can also simplify 
the software implementation of the resulting solution and/or facilitate a more rigorous characterization of 
its performance. We begin in the next section by providing broader contextual background 
and then proceeding to examine a concrete example for which a conjectured property is
formally established. The proven property is not as important in and of itself as it is an
illustrative example of a kind of property that might be identified and speculated about in a given
application but not actually investigated more deeply because it is not amenable to any standard form of 
pencil-and-paper analysis. As a byproduct, we use the example problem to demonstrate
a novel dimension-pinning strategy that is likely to be relevant to a broad spectrum of practical problems.

\section{Background}

This paper is in part inspired by recently-revived discussions
about the current and potential prevalence of 
computer-assisted proof methods spurred by the
quarter-century anniversary of the notorious 
``Death of Proof'' {\em Scientific American} article from 1993\,\cite{dop1}.
While that article predicted a massive rise in the power and widespread use
of automated-proof methods -- {\em with a  commensurate decline in the
role of human mathematicians} --  its 2019 follow-up,
``Okay, Maybe Proofs Aren't Dying After All,'' concludes that 
evidence now tends to undercut the thesis of the original
paper\,\cite{dop2}. The author's pessimistic reassessment was motivated
by observations such as: ``{\em...computers continue to only rarely 
have any role in the creation and checking of the proofs of 
mathematical theorems}''\,\cite{dop2}. Conventional current
opinion is that this status quo will not change unless 
and until the power of automated proof methods 
improves dramatically to a point where they are able to solve
an increasing fraction of outstanding problems of interest
to professional mathematicians. In this paper, however,
we suggest an alternative in which their power may increase 
at only a moderate rate but there is a dramatic increase in
the number of problems to which they are applied.
These problems are most likely to found in applied mathematics and engineering
applications and are of only limited theoretical interest to the mathematical 
community because of their highly-specialized practical applications (e.g., as typified 
by the example problem of this paper) but which by sheer number lead to a 
situation in which optimization-based automated methods do in fact surpass the productivity and 
practical impact of traditional proof methods employed by human mathematicians.

As an illustrative example, we provide a case study demonstrating
the application of sum-of-squares (SOS) optimization methods 
to a problem involving
the establishing of special properties that hold only in a fixed
number of dimensions. Unlike results that hold for an infinite
domain of objects such as the real or complex numbers -- which tend
to be amenable to relatively succinct human-derivable proofs -- 
the restriction to dimension-specific properties typically leads
to proofs with complexity that grows exponentially with dimension.
It must be emphasized that our example problem is specifically intended 
to show the kinds of highly specialized results that can potentially
be proven in the context of a given practical application. In other
words, the results that are proven are not primarily 
of significance in and of themselves but rather serve to demonstrate
the extent to which a vast number of similarly specialized results 
can be expected to hold in a wide variety of applications but 
typically go unrecognized. In other words, we argue that many 
engineering system designs and implementations fail to exploit 
specialized properties of their problem domain.

\section{Dimension-Specific Properties}

For a particular class of objects of interest, e.g., some
defined subset of the integers or reals, it is commonly
of analytic or practical value to identify a set of special
properties satisfied by that class. For a class of objects
parameterized by an integral measure of size or
dimensionality $d$, it may be the case that a particular
property only holds for a limited set of values of $d$, e.g.,
$d<7$.  Properties relating to a system of interest 
that are found to hold only up to a fixed number of 
dimensions (i.e., degrees of freedom) tend to be both 
surprising and revealing. More importantly, they can 
potentially be of significant practical value for applications 
that fall within that span of dimensionality. Unfortunately,
properties of this kind tend to be difficult to prove.

In the case of $n\times n$ matrices there is great
value in being able to establish that the result of a
given matrix function or transformation has a
special property (e.g., that it is unitary, positive-definite,
totally-unimodular, doubly-stochastic, or has a special
structural property such as sparseness) which can
be exploited for analysis purposes or to obtain solutions
more efficiently than would be possible in the general
case. This is the motivation for deriving decompositions
of general matrices as products or sums of matrices with
special properties\,\cite{mathmag,mil,varga}. For example,
the singular-value decomposition (SVD) is one of the most
widely-used tools in linear algebra because it permits an
arbitrary matrix $M$ to be expressed as $M=UDV^*$,
where $U$ and $V$ are unitary and $D$ is diagonal\,\cite{hj1}.

When general linear-algebraic tools are applied to analyze
a matrix transformation $f(M)$, the identified properties
will typically hold independently of $n$.
In fact, the power of linear algebra largely derives from its
ability to work with objects (matrices) in a manner that
allows the dimensionality to be abstracted out of the
formulation of a given problem. However, many
theoretical and practical problems of interest, e.g.,
design and analysis of control systems,
are intrinsically defined in a fixed $d$-dimensional space for
which many properties may hold for $f(M)$ conditioned
on $n\leq d$ that do not necessarily hold for general $n$.
Once identified, such properties can provide significant
insights and permit a much larger set of mathematical tools
to be applied. Unfortunately, establishing such properties
may demand effort exponential in dimensionality.

In the following sections we consider an example in 
which the opportunities and obstacles associated with
the determination of dimension-specific properties
are highlighted. In particular, we prove conjectured
properties of a specific $n\times n$ matrix function
for $n\leq 4$ using SOS optimization methods for
certifying polynomial nonnegativity. We then
consider the challenge of proving that these
properties hold more generally up to a conjectured
upper bound of $n=6$. We conclude with
consideration of {\em dimension pinning}
as a potentially effective and general strategy for
establishing dimension-specific properties as part of
a broader vision for highlighting the applicability and
relevance of computer-assisted proof methods to
applied mathematics and engineering.

\section{The RGA and its Inverse}

The relative gain array (RGA) is an important tool in the
design of process control systems\,\cite{bristol}.
It is a function of a nonsingular real or
complex\footnote{It should be noted that the 
transpose operator in the RGA applies even in
the case of complex $G$, i.e., it should {\em not}
be replaced with the {\em conjugate}-transpose
operator.}
matrix $G$ defined as:
\begin{equation}
   \RGA(G) ~\doteq~ G \circ (G^{-1})^T 
   ~\equiv~ G \circ G^{-T}
\end{equation}
where $\circ$ represents the elementwise Hadamard matrix
product and the rightmost expression exploits
simplified notation for the inverse-transpose operator. 
The RGA has a variety of interesting mathematical properties,
one of which is that the sum of the elements in each row and
column is unity\,\cite{rgajs,ucrga}. For example, given\footnote{This
particular integer matrix was chosen purely because its inverse
is also integral\,\cite{hanson} and thus simplifies the values
appearing in this and subsequent examples, i.e., no inferences
should be drawn from the integrality of the results.}
\begin{equation}
   \label{simpmat}
   G ~=\, \left[\!\!\begin{array}{rrr}
                        1 & -2 & 3\\
                        2 & -5 & 10\\
                        -1 & 2 & -2
                  \end{array}\right]
\end{equation}
then
\begin{equation}
   \RGA(G) ~=\, \left[\!\begin{array}{rrr}
                        10 & -12 & 3\\
                        -4 & 5 & 0\\
                        -5 & 8 & -2
                  \end{array}\right] .
\end{equation}
Another important property of the RGA relates to
permutations of the rows and columns of its argument \cite{rgajs}:
\begin{equation}
   \RGA(P\cdot G\cdot Q) ~=\, P\cdot\RGA(G)\cdot Q.
\end{equation} 
In other words, the RGA is {\em permutation-consistent} \cite{usiam,ucrga}
with respect to left and right multiplication by permutation matrices $P$ and $Q$.

A related matrix operator that has not previously been
considered in the literature is the {\em inverse} of the
RGA result, or IRGA, which we define as:
\begin{equation}
   \IRGA(G) ~\doteq~ \left(G \circ G^{-T} \right)^{-1}.
\end{equation}
Applying this to the matrix of Eq.\,(\ref{simpmat}) gives
\begin{equation}
   \IRGA(G) ~=\,\frac{1}{25} \left[\!\!\begin{array}{rrr}
                        10 & 0 & 15\\
                        8 & 5 & 12\\
                        7 & 20 & -2
                  \end{array}\right]
\end{equation}
where it can be seen that the row and column sums
are unity, as should be expected since $\RGA(G)$
has unit row and column sums. Examining the
$2\times 2$ case
\begin{equation}
   G  ~=\,\left[\!\begin{array}{rr}
                        a & b\\
                        c & d
                  \end{array}\right]
\end{equation}
reveals
\begin{equation}
  \IRGA(G)  ~=\,\frac{1}{ad+bc} \left[\!\begin{array}{rr}
                        ad & bc\\
                        bc & ad
                  \end{array}\right].
\end{equation}
This result suggests several special properties that might be
conjectured to hold for some values $n>2$. The most tantalizing
relates to the fact that the denominator of the coefficient is the
{\em permanent} of $G$. In fact, the overall result represents the
joint assignment matrix (JAM) \cite{jcju} of $G$, which is the 
solution\footnote{The inverse-JAM is presently
under investigation by the first author as an alternative generalization
of the RGA for $d>2$.}
to an important combinatorial problem arising in multiple-target
tracking and related applications\,\cite{franklin,crc11}.
Evaluation of the JAM for a general $n\times n$ matrix is believed
to be computationally intractable based on the \#P-Hard complexity of
evaluating the permanent of a matrix\,\cite{jcju,atanasov,moreland,crouse},
so it should not be surprising that JAM-equivalence does not hold for $n>2$.

It can also be observed in the $2\times 2$ case that
nonnegative $G$ implies $\IRGA(G)$ is also
nonnegative and therefore doubly-stochastic.
Unfortunately, this property also does not
generalize to $n>2$. However, in the case
of positive-definite $G$ a conjecture of the
first author states:
\begin{conjecture}
For an $n\times n$, $n\leq 6$,
positive-definite matrix $G$, $\IRGA(G)$ is
positive-definite, has unit row and column sums, and
is nonnegative and hence doubly-stochastic.
\end{conjecture}
In fact, because of the permutation-consistency
property of the RGA, hence also of the IRGA, the conjecture 
only requires $G$ to be positive-definite up to left and right
permutations.

A nonnegative symmetric positive-definite matrix is
sometimes referred to as being {\em doubly
nonnegative}\,\cite{doublynonneg}. Such matrices
arise in a variety of applications ranging from
control systems and network analysis to estimation
and optimization\,\cite{altafini,pohv,rahmani,vosughi,zhu},
and characterizations of their inverses have been
long-studied\,\cite{fiedler1,fiedler2,royxue}.

It is straightforward to prove that positive-definiteness
of the IRGA is preserved for all $n$ because
$G$ and its inverse are both PD (positive-definite), and the Hadamard
product of two PD matrices always yields a PD
result\,\cite{hj1,schur}. Therefore the critical property to be
verified for values of $n$ in the conjectured
range is nonnegativity. This is the central example property
of interest in this paper for motivating the broad relevance
of automated optimization-based proof methods.
To this end, while the details of the solution method of the following
section are illustrative, the principal
takeaway is that the optimization-based 
certification of polynomial nonnegativity can actually yield
a proof that the IRGA is in fact nonnegative for $d\leq 4$
and provide compelling evidence that the result holds
up to the conjectured $d\leq 6$.

\section{Certifying nonnegativity of polynomials}

Let $\R[\x]=\R[x_1,\ldots,x_n]$ be the ring of real $n$-variate polynomials. For a finite set $\A\subset\N^n$, we denote by $\Conv(\A)$ the convex hull of $\A$. A polynomial $f\in\R[\x]$ can be written as $f(\x)=\sum_{\a\in\A}c_{\a}\x^{\a}$ with $c_{\a}\in\R, \x^{\a}=x_1^{\alpha_1}\cdots x_n^{\alpha_n}$. The support of $f$ is $\supp(f):=\{\a\in\A\mid c_{\a}\ne0\}$.

For a nonempty finite set $\B\subseteq\N^n$, $\R[\B]$ denotes the set of polynomials in $\R[\x]$ whose supports are contained in $\B$, i.e., $\R[\B]=\{f\in\R[\x]\mid\supp(f)\subseteq\B\}$ and we use $\R[\mathscr{B}]^2$ to denote the set of polynomials which are sums of squares of polynomials in $\R[\B]$. The set of $r\times r$ symmetric matrices is denoted by $S^r$ and the set of $r\times r$ positive-semidefinite matrices is denoted by $S_+^r$. Let $\x^{\mathscr{B}}$ be the $|\mathscr{B}|$-dimensional column vector consisting of elements $\x^{\b},\b\in\mathscr{B}$, then
\begin{equation*}
\R[\mathscr{B}]^2=\{(\x^{\mathscr{B}})^TQ\x^{\mathscr{B}}\mid Q\in S_+^{|\B|}\},
\end{equation*}
where the matrix $Q$ is called the {\em Gram matrix}.

A classical approach for checking nonnegativity of multivariate polynomials, as introduced by Lasserre\,\cite{la} and Parrilo\,\cite{pa}, is the use of sums of squares as a suitable replacement for nonnegativity. Given a polynomial $f(\x)\in\R[\x]$, if there exist polynomials $f_1(\x),\ldots,f_m(\x)$ such that
\begin{equation}\label{sec1-eq1}
f(\x)=\sum_{i=1}^mf_i(\x)^2,
\end{equation}
then $f(\x)$ is referred to as a {\em sum of squares} (SOS). Obviously an SOS decomposition of a given polynomial serves as a certificate for its nonnegativity. For $d\in\N$, let $\N^n_d:=\{\a\in\N^n\mid\sum_{i=1}^n\alpha_i\le d\}$ and assume $f\in\R[\N^n_{2d}]$. Then the SOS condition (\ref{sec1-eq1}) can be converted to the problem of deciding if there exists a positive-semidefinite matrix $Q$ such that
\begin{equation}\label{sec1-eq2}
f(\x)=(\x^{\N^n_{d}})^TQ\x^{\N^n_{d}},
\end{equation}
which can be effectively solved as a semidefinite programming (SDP) problem. Note that there are $\binom{n+d}{n}$ elements in the monomial basis $\x^{\N^n_{d}}$. So the size of the corresponding SDP problem is $O(\binom{n+d}{n})$, which grows rapidly as the number of variables $n$ and the degree $2d$ of the given polynomial increase. To deal with high-degree polynomials with many variables it is crucial to exploit the sparsity of polynomials to reduce the size of the corresponding SDP problems.

\section{Certifying nonnegativity of sparse polynomials}
There are two aspects by which sparsity in SOS decompositions can be exploited. The first is to compute a smaller monomial basis. In fact, the monomial basis $\N^n_{d}$ in (\ref{sec1-eq2}) can be replaced by \cite{re}
\begin{equation}\label{sec2-eq1}
\B=\Conv(\{\frac{\a}{2}\mid\a\in\supp(f)\})\cap\N^n\subseteq\N^n_{d}.
\end{equation}
Second, the block-diagonal structure in the Gram matrix $Q$ can be exploited, namely {\em cross sparsity patterns}, which was introduced by the second author\,\cite{wang} and will prove critical to achieving our main result.
\begin{definition}
Let $f(\x)\in\R[\x]$ with $\supp(f)=\mathscr{A}$ and $\B=\{\bo_1,\ldots,\bo_r\}$ is as in (\ref{sec2-eq1}). An $r\times r$ {\em cross sparsity pattern matrix} $\mathbf{R}_{\mathscr{A}}=(R_{ij})$ is defined by
\begin{equation}\label{sec2-eq2}
R_{ij}=\begin{cases}
1, &\bo_i+\bo_j\in\mathscr{A}\cup(2\N)^n,\\
0, &\textrm{otherwise}.
\end{cases}
\end{equation}
Given a cross sparsity pattern matrix $\mathbf{R}_\mathscr{A}=(R_{ij})$, the graph $G(V_\mathscr{A},E_\mathscr{A})$ where $V_\mathscr{A}=\{1,2,\ldots,r\}$ and $E_\mathscr{A}=\{(i,j)\mid i,j\in V_\mathscr{A}, i<j, R_{ij}=1\}$ is called the {\em cross sparsity pattern graph}.
\end{definition}

Given an undirected graph $G(V,E)$ with $V=\{1,2,\ldots,r\}$, we define an extended set of edges $E^{\star}:=E\cup\{(i,i)\mid i\in V\}$ that includes all self-loops. Then we define the space of symmetric sparse matrices as
\begin{equation}\label{sec2-eq3}
S^r(E,0):=\{X\in S^r\mid X_{ij}=X_{ji}=0\textrm{ if }(i,j)\notin E^{\star}\}
\end{equation}
and the cone of sparse positive-semidefinite matrices as
\begin{equation}\label{sec2-eq4}
S^r_+(E,0):=\{X\in S^r(E,0)\mid X\succeq0\}.
\end{equation}

Assume $\A\subseteq\N^n$ and $\B=\{\bo_1,\ldots,\bo_r\}$ is as in (\ref{sec2-eq1}). Let $G(V_\mathscr{A},E_\mathscr{A})$ be the cross sparsity pattern graph. Let $\widetilde{G}(V_\mathscr{A},\widetilde{E}_\mathscr{A})$ be the graph obtained by adding edges to all connected components of $G(V_\mathscr{A},E_\mathscr{A})$ such that every connected component becomes a complete subgraph. Letting
\begin{eqnarray*}
\widetilde{\Sigma}(\mathscr{A}) & :=&\{f\in\R[\A]\mid\exists Q\in S_+^r(\widetilde{E}_{\mathscr{A}},0)\\ 
~ & ~ & \hspace{1.9cm}\textrm{ s.t. }f=(\x^{\mathscr{B}})^TQ\x^{\mathscr{B}}\},
\end{eqnarray*}
we can state the following sparse SOS decomposition theorem:
\begin{theorem}[\cite{wang}]\label{sec2-thm1}
Assume $\A\subseteq\N^n$ and $\B=\{\bo_1,\ldots,\bo_r\}$ is as in (\ref{sec2-eq1}). Let $C_1, C_2, \ldots, C_t\subseteq V_\mathscr{A}$ denote the connected components of $\widetilde{G}(V_\mathscr{A},\widetilde{E}_\mathscr{A})$ and $\mathscr{B}_k=\{\bo_i\in\B\mid i\in C_k\}, k=1,2,\ldots,t$. Then, $f(\x)\in\widetilde{\Sigma}(\mathscr{A})$ if and only if there exist $f_k(\x)\in\R[\mathscr{B}_k]^2$ for $k=1,\ldots,t$ such that
\begin{equation}\label{sec2-eq5}
f(\x)=\sum_{k=1}^tf_k(\x).
\end{equation}
\end{theorem}

Theorem \ref{sec2-thm1} implies that the checking of $f(\x)\in\widetilde{\Sigma}(\mathscr{A})$ involves solving a block-diagonalized SDP problem, which significantly reduces the computation. For proof purposes it is then necessary to convert the obtained numerical SOS decomposition to an SOS decomposition with rational coefficients. Standard rounding-projection procedures \cite{pe} can be applied
when the given polynomial lies in the interior of the SOS cone, but in our case $J_{412}$ (and also $J_{512}$, $J_{612}$) lies on the boundary of the SOS cone. To overcome this difficulty we use the method of undetermined coefficients by setting nonzero elements in the numerical SOS decomposition of $J_{412}$ as unknowns and then searching for a rational solution to the system of equations obtained by comparing coefficients. It is the complexity of this latter step that will limit the values of $n$ for which SOS certificates can be practically obtained for our problem of interest.

\section{Toward Proving the Conjecture for $n\leq 6$}

The basis for the upper bound of the conjecture was
determined first by a strategy of extensive directed sampling
of random PD matrices for successive values of $n$, which
revealed that counterexamples can be readily found for
$n=7$ but not for $n=6$. It then remained only to
tailor the $n=7$ search to find a specific example for which the
violated nonnegativity condition can be certified with
finite arithmetic. The following integer PD matrix provides this
certification:
\begin{equation}
 \left[\!\begin{array}{rrrrrrr}
   54 & -10 &   0 &   0 &  -4 & -2 &  -4\\
  -10 &  55 &  -5 &  -3 &  -3 &  1 &  -1\\
    0 &  -5  & 52  & -8  &  2  & -2  & -4\\
    0 &  -3  & -8  & 57  & -7  & -3  & -1\\
   -4 &  -3  &  2  & -7   &52  & -6  &  1\\
   -2 &   1  & -2  & -3   &-6  & 56  & -9\\
   -4 &  -1  & -4  & -1   & 1  & -9  & 53
                  \end{array}\,\right]
\end{equation}
Applying IRGA to the above can be verified to
yield a rational result that contains a symmetric pair
of negative values. With this it remains to
rigorously prove that no such counterexamples
can exist for $n=6$. This is necessary and sufficient
because clearly if the conjecture holds for some
value of $n$ it must hold for all smaller values
because each case can be expressed as a block
replacing a submatrix of the $n\times n$
identity.

To prove that the $6\times 6$ case is nonnegative
it is sufficient to show that a typical off-diagonal
element is nonnegative. As will be discussed in the
subsequent section, the computational complexity
required to prove the nonnegativity of the polynomial
associated with such an element tends to become
practically prohibitive even for values of $n$ of
this size.

\section{The $6\times6$ case}
Any positive-definite matrix can be be scaled on the left and right by a positive diagonal matrix so that its Cholesky decomposition has unit diagonal elements, thus we can assume without loss of generality that our test of
positive-definiteness can assume an appropriately-scaled $\G\in\R^{6\times6}$ having Cholesky factors $\G=LL^T$ with
\begin{equation}
L ~=~ 
 \left[\,\begin{array}{cccccc}
1&0&0&0&0&0\\
a&1&0&0&0&0\\
b&c&1&0&0&0\\
d&e&f&1&0&0\\
g&h&i&j&1&0\\
k&l&m&n&p&1
\end{array}\,\right]
\end{equation}
Because $\G\circ (\G^{-1})^T$ is positive-definite it follows that the determinant of $\G\circ (\G^{-1})^T$, which is also the common denominator of elements of $\IRGA(\G)=(\G\circ(\G^{-1})^T)^{-1}$, is positive. Therefore, to prove the nonnegativity of elements of $\IRGA(\G)$ it is sufficient to establish nonnegativity of the numerators of elements of $\IRGA(\G)$. Consider a typical off-diagonal element of $\IRGA(\G)$, say the element in position $(1,2)$, whose numerator we denote by $J_{612}$ (similarly for the meaning of $J_{412},J_{512}$). The sizes of $J_{412},J_{512},J_{612}$ are listed in the following table from which it can be seen that the support sizes of $J_{512},J_{612}$ become extremely large:

\begin{table}[htbp]
  \begin{center}
    \begin{tabular}{|c|c|c|c|c|}
      \hline
       &\#var&deg&\#supp    \\
      \hline
      $J_{412}$ & 6 & 12& 116\\
      \hline
      $J_{512}$ & 10 & 20& 5157\\
      \hline
       $J_{612}$ & 15 & 40& 676505\\
      \hline
    \end{tabular}
  \end{center}
  \caption{Sizes of $J_{412}$,$J_{512}$,$J_{612}$.}
\end{table}

Applying the method of the previous section we obtain a numerical SOS decomposition for $J_{512}$,
but its size prevents us from obtaining a rational certificate of nonnegativity. In other words,
IRGA nonnegativity for $n=5$ is virtually but not formally established. However, the much larger
size of $J_{612}$ prevents us from obtaining even a numerical SOS decomposition, assuming
one exists.

Given that it is not presently possible to formally
prove nonnegativity for the $n=5$ case,
the fallback must be to follow a strategy
similar to the finding of counterexamples
and consider successive values of $n>2$.
The $n=3$ case is already at the limit of
what can be proven by hand or by using
computer-aided symbolic methods (both are
doable), but in the following section we
will formally prove nonnegativity for the
subsuming case of $n=4$, i.e., that
typical element $J_{412}$ is nonnegative.

\section{The $4\times4$ case}

As in the previous section, assume the positive-definite matrix $\G\in\R^{4\times4}$ has Cholesky decomposition $\G=LL^T$ with
\begin{equation}
L ~=~ 
 \left[\,\begin{array}{cccc}
1&0&0&0\\
a&1&0&0\\
b&c&1&0\\
d&e&f&1
\end{array}\,\right]
\end{equation}

From this we can express the $J_{412}$ polynomial as
\begin{align*}
J_{412}=&-(d^2+e^2+f^2+1) (a (f^2+1) (b^2\\
&+c^2+1) (-a c^2 f^2-a c^2+2 a c e f\\
&-a e^2-a+b c f^2+b c-b e f-c d f+d e)\\
&-b (a b+c)(-c f^2-c+e f)\\
&(a c f^2+a c-a e f-b f^2-b+d f))\\
&-(a d+e)(c f-e)(d (f^2+1)(-(b^2\\
&+c^2+1))(-a c f+a e+b f-d)\\
&-b f (b d+c e+f) (a c f^2+a c-a e f\\
&-b f^2-b+d f))-f (b d+c e+f)(d (-(a b\\
&+c))(-c f^2 -c+e f)(-a c f +a e+b f-d)\\
&-a f (b d+c e +f)(-a c^2 f^2-a c^2+2 a c e f\\
&-a e^2-a+b c f^2+b c-b e f-c d f+d e)).
\end{align*}
Computing the monomial basis of $J_{412}$ as in (\ref{sec2-eq1}) yields a result with $48$ elements. The cross sparsity pattern graph of $J_{412}$ has $10$ connected components with $12,10,7,7,4,3,2,1,$ $1,1$ nodes, respectively. From this we obtain $J_{412}$ as a weighted sum of $25$ squares, i.e., $J_{412}=\sum_{i=1}^{25}p_i$, where
\allowdisplaybreaks
\begin{align*}
     p_1= &~ 2 (a c d f-a b e f)^2,\\
     p_2=&~\mbox{\large (}\frac{1}{2} a f^2 c^2+\frac{a c^2}{2}-\frac{1}{2} b f^2 c+\frac{d f c}{2}\\
            &~ -a e f c-\frac{b c}{2}+\frac{a e^2}{2}+a+\frac{b e f}{2}-\frac{d e}{2}\mbox{\large )}^2,\\
     p_3=&~\frac{3}{4} \mbox{\large (}\frac{1}{3} a f^2 c^2+\frac{a c^2}{3}-\frac{1}{3} b f^2 c+\frac{2}{3} a e f c\\
            &~~ -\frac{b c}{3}-\frac{d f c}{3}-a e^2+d e-\frac{b e f}{3}\mbox{\large )}^2,\\
     p_4=&\frac{2}{3}\! \left(\!\! -a f^2\! c^2\!-a c^2\! +b f^2 c+b c+a e f c-\!\frac{d f c}{2}-\!\frac{b e f}{2}\right)^2\!\!\! ,\\
     p_5=&~\frac{1}{2} \left(a c^2 f^2-b c f^2+c d f-a c e f\right)^2,\\
     p_6=&~\frac{1}{2} \left(a c^2 f^2-b c f^2+b e f-a c e f\right)^2,\\
     p_7=&~\mbox{\large (}\! -\frac{1}{2} a e f c^2+\frac{1}{2} a e^2 c\\
            &~~ -\frac{1}{2} a f^2 c+a c+\frac{1}{2} b e f c-\frac{d e c}{2}+\frac{a e f}{2}\mbox{\large )}^2,\\
     p_8=~&\frac{3}{4} \left(a e f c^2\!-a e^2 c-\frac{1}{3} a f^2 c+d e c-b e f c+\!\frac{a e f}{3}\right)^2\!\! ,\\
     p_9=~&\frac{2}{3} \left(a e f-a c f^2\right)^2,\\
     p_{10}=&\left(-\frac{1}{2} a d f c^2+\frac{1}{2} b d f c+\frac{1}{2} a b e^2+a b-\frac{b d e}{2}\right)^2,\\
     p_{11}=&~\frac{3}{4} \left(a d f c^2-b d f c-a b e^2+b d e\right)^2,\\
     p_{12}=&\left(a d f-a b f^2\right)^2,\\
     p_{13}=&\left(\frac{1}{2} a e c^2-\frac{b e c}{2}-\frac{a f c}{2}+a e\right)^2,\\
     p_{14}=&\frac{3}{4} \left(-a e c^2+b e c-a f c\right)^2,\\
     p_{15}=&~ 2 a^2 c^2 f^2,\\
     p_{16}=&\left(\frac{1}{2} a d c^2-\frac{b d c}{2}+a d\right)^2,\\
     p_{17}=&~ \frac{3}{4} \left(-a d c^2+b d c-\frac{2 a b f}{3}\right)^2,\\
     p_{18}=&~ \frac{5}{3} a^2 b^2 f^2,\\
     p_{19}=&~ a^2 c^2 d^2,\\
     p_{20}=&~2 \mbox{\large (}\frac{1}{2} a c^2 f^3-\frac{1}{2} b c f^3+\frac{1}{2} c d f^2+\frac{1}{2} b e f^2-a c e f^2\\
                &~~~ +\frac{1}{2} a c^2 f+\frac{1}{2} a e^2 f+a f-\frac{b c f}{2}-\frac{d e f}{2}\mbox{\large )}^2,\\
     p_{21}=&~2 \mbox{\large (}\!\! -\frac{1}{2} a c^2 f^3+\frac{1}{2} b c f^3-\frac{1}{2} c d f^2-\frac{1}{2} b e f^2\\
                &~~~~ +a c e f^2-\frac{1}{2} a e^2 f+\frac{d e f}{2}\mbox{\large )}^2,\\
     p_{22}=&~ \frac{1}{2} \left(b c f-a c^2 f\right)^2,\\
     p_{23}=&~ a^2 c^2 e^2,\\
     p_{24}=&~ a^2 b^2 d^2,\\
     p_{25}=&~ a^2 b^2 e^2.
\end{align*}

We can now present our main
technical result relating to the IRGA conjecture:
\begin{theorem}
For an $n\times n$ matrix $G$ that is positive-definite
up to permutations of its rows and columns, $\IRGA(G)$ 
retains the same properties, but with unit row and column sums,
and for $n\leq 4$ is nonnegative and hence doubly-stochastic, i.e.,
is doubly-nonnegative.
\end{theorem}
The import of this theorem is that for
any application of the IRGA to arbitrarily permuted
positive-definite $n\times n$ matrices, $n\leq 4$,
it can be rigorously assumed that the result will 
retain the same properties while also being
doubly-stochastic.

\begin{remark}
Obtaining an SOS decomposition with rational coefficients for $J_{412}$ required finding a rational solution to a system of $19$ polynomial equations of degree $3$ and with $28$ unknowns. The corresponding system for $J_{512}$ involves many hundreds of unknowns and is presently beyond the limits of what can be solved using existing generic methods. However, we believe that a specially-tailored approach may be able to obtain a rational SOS decomposition in the near future and formally establish the nonnegativity of $J_{512}$. Obtaining a solution for $J_{612}$ may require completely new theoretical and practical innovations, e.g., possibly exploiting other special properties of PD doubly-stochastic matrices that obviate the need for SOS methods or greatly reduce the complexity of the polynomials to which they must be applied. With or without such innovations, we propose $J_{612}$ as a challenge problem to the SOS community.
\end{remark}

It must be emphasized that the focus of this paper is not on
this highly-specific problem/result {\em per se} but rather 
how it serves as an example of the kinds of highly-specialized
properties of relevance to practical applications that
can potentially be established using optimization-based proof 
methods. Such
methods go well beyond the polynomial-nonnegativity tools
used in this paper and of course include general-purpose 
mathematical theorem provers. What is argued in this paper
is that a closer look will reveal a plethora of novel
application-specific properties that can be practically
established/proven using present state-of-the-art tools.

\section{Discussion}

It can be argued that progress in mathematics has been
biased toward types of problems that tend to lead to succinctly
stated theorems which admit comparably succinct proofs.  
Although the four-color conjecture for planar maps arose from this tradition, 
the only known proofs for the celebrated four-color {\em theorem}  
appear to demand computer-assisted methods to resolve the status of hundreds 
of complicated special-case subproblems \cite{appel1,appel2,appel3,robertson}.
The nature of the original 1977 proof \cite{appel1,appel2} was considered surprising 
and rather inelegant because most related problems, e.g., the
looser five-color variant, admitted relatively succinct
conventional proofs in \cite{heawood,robertson}. 
A natural reaction to the daunting complexity of the four-color proof is to 
regard it as an ``outlier'' or ``special case,'' e.g., like early
Greek mathematicians may have regarded their first-discovered
examples of irrational numbers. 

In this paper we have argued that the class of practical
problems for which optimization-based proof methods are
required is vastly larger than currently recognized by the computer science
community and includes
many areas of applied mathematics and engineering. We have
discussed one such problem class, which involves the establishing
of dimension-specific properties, and we have provided
a case study to serve as a representative example. 
In particular, we examined a problem for
which it is possible to identify and prove nontrivial
properties relating to a strongly nonlinear
$n\times n$ matrix function that hold for values
(dimensionality) of $n$ below a specific bound but
do not hold for general $n$ above a specific bound.
Such computer-established properties can be exploited
not only during system design but also to facilitate
system implementation and testing. Our dimension-pinning
approach, which is likely to be applicable to
a broad family of problems, uses an automated
counterexample search to establish an upper-bound limit 
on $n$, i.e., above which the speculated properties provably 
do not generally hold, and then uses a semi-automated
optimization-based proof method 
to establish that the speculated properties hold in general 
below that bound. Ideally, the bounds obtained from the
two steps would converge to
optimally define/pin the limit at which the properties fail
to hold in all cases.

Of course there are myriad examples dating from the
earliest history of mathematics in which explicit counterexamples
have been used to demonstrate the limit to which the
scope of a given result can be generalized. What we
advocate here is a more structured approach to
facilitate computer-aided proof methods to be
applied to identify and establish potentially valuable
properties of mathematical formulae that hold only
in low dimensions, e.g., sufficient to encompass most or all
practical instances of a particular engineering problem.
For the specific problem of this paper, the nonnegativity
of the IRGA of a positive-definite matrix was observed
for the case of $n=2$, and this motivated a counterexample
search which identified that nonnegativity does not
generally hold for $n\geq 7$. This then motivated
use of SOS methods to prove that nonnegativity does hold
in general at least up to $n=4$. SOS methods have also
provided strong evidence that nonnegativity holds for
$n=5$, but due to computational complexity constraints
inherent to SOS methods the status of $n=6$
presently remains unresolved -- aside from strong
supporting evidence from the
fact that extensive search has yielded no counterexamples
for $n=6$ whereas they are readily found when $n=7$.

We have noted that dimensionality-limited properties are
particularly relevant to those applications for
which $n$ is practically limited in some way by the physical dimensions
of the real world. Beyond this, a proof of the nonnegativity
conjecture of this paper would imply the existence of special
mathematical structures that are limited to 6-dimensional
spaces and thus may be relevant to theoretical physics
due to the fact that unitary conformal field theories (CFTs) and superconformal
variants can only be defined in $n=4$ or $n=6$
dimensions\,\cite{witten1,witten2,aharony,saemann,samtleben}.
Most importantly, we hope this paper has provided a glimpse
of what we believe to be a diverse and bountiful -- {\em though
heretofore largely unrecognized} -- set of problems to
which optimization-based automated proof methods can be productively
applied.

\vspace{11pt}
\noindent {\bf Declarations of Interest}: None.
\vspace{6pt}

\noindent {\bf Acknowledgements}: The first author wishes to thank
Charlie Johnson (\cite{hj1,rgajs}) for enjoyable
conversations relating to the topic of this paper.
The second author wishes to gratefully acknowledge support from 
the China Postdoctoral Science Foundation under grants 2018M641055.

\bibliographystyle{amsplain}

\end{document}